\documentclass[apjl]{emulateapj}
\usepackage{color}

\usepackage{graphicx}

\begin{document}

\title{Thermal Infrared Imaging of MWC 758 with the Large Binocular Telescope: Planetary Driven Spiral Arms?}

\color{black}
\shorttitle{LBT Imaging of MWC 758}
\shortauthors{Wagner et al.}
\author{Kevin Wagner\altaffilmark{1,2}$^{\star}$, Jordan M. Stone\altaffilmark{1,3}, Eckhart Spalding\altaffilmark{1}, Daniel Apai\altaffilmark{1,2,4}, Ruobing Dong\altaffilmark{5}, Steve Ertel\altaffilmark{6,1}, Jarron Leisenring\altaffilmark{1}, \& Ryan Webster\altaffilmark{1}}

\altaffiltext{1}{Steward Observatory, University of Arizona}
\altaffiltext{2}{NASA NExSS \textit{Earths in Other Solar Systems} Team}
\altaffiltext{3}{NASA Hubble Postdoctoral Fellow}
\altaffiltext{4}{Lunar and Planetary Laboratory, University of Arizona}
\altaffiltext{5}{University of Victoria, British Columbia, Canada}
\altaffiltext{6}{Large Binocular Telescope Observatory, University of Arizona}

\altaffiltext{$\star$}{Correspondence to: kwagner@as.arizona.edu}

\keywords{Stars: pre-main sequence (MWC 758)  --- planets and satellites: formation --- planets and satellites: detection --- planet-disk interactions }

\begin{abstract}
Theoretical studies suggest that a giant planet around the young star MWC 758 could be responsible for driving the spiral features in its circumstellar disk. Here, we present a deep imaging campaign with the Large Binocular Telescope with the primary goal of imaging the predicted planet. We present images of the disk in two epochs in the $L^{\prime}$ filter (3.8 $\mu m$) and a third epoch in the $M^{\prime}$ filter (4.8 $\mu m$). The two prominent spiral arms are detected in each observation, which constitute the first images of the disk at $M^\prime$, and the deepest yet in $L^\prime$ ($\Delta L^\prime=$12.1 exterior to the disk at 5$\sigma$ significance). We report the detection of a S/N$\sim$3.9 source near the end of the Sourthern arm, and, from the source's detection at a consistent position and brightness during multiple epochs, we establish a $\sim$90\% confidence-level that the source is of astrophysical origin. We discuss the possibilities that this feature may be a) an unresolved disk feature, and b) a giant planet responsible for the spiral arms, with several arguments pointing in favor of the latter scenario. We present additional detection limits on companions exterior to the spiral arms, which suggest that a $\lesssim$4 M$_{Jup}$ planet exterior to the spiral arms could have escaped detection. Finally, we do not detect the companion candidate interior to the spiral arms reported recently by \cite{Reggiani2018}, although forward modelling suggests that such a source would have likely been detected.
\end{abstract}


\section{Introduction}

Recent images of protoplanetary disks have indicated that substructures are common, and that planet-disk interactions are likely to be on-going in a number of young systems (e.g., \citealt{Muto2012}, \citealt{Quanz2013a}, \citealt{Debes2016}, \citealt{Andrews2018}, \citealt{Zhang2018}). Commonly observed substructures include gaps and spiral arms, among others that may be linked to planet formation (e.g., \citealt{Zhu2012}, \citealt{Dong2015a}, \citealt{Dong2015b}). Meanwhile, high-contrast visible to thermal infrared imaging has enabled the direct detection of young and forming exoplanets (e.g., \citealt{Keppler2018}, \citealt{Wagner2018a}). Combined, these techniques now allow for giant planet formation to be studied in unprecedented detail, as the first causal connections between observed planet properties and protoplanetary disk substructures are being established (e.g., \citealt{Quanz2013b}, \citealt{Apai2015}, \citealt{Muley2019}).


Spiral features represent one class of disk substructures that are frequently hypothesized to be related to planet formation. Spirals can arise from gravitational instability of massive disks (\citealt{Kratter2016}), from time-variable disk shadowing (\citealt{Montesinos2016}), or from the dynamical effects of an orbiting companion (e.g., \citealt{Fung2015}). Most disks do not appear massive enough to drive two-armed spirals through gravitational instability (\citealt{Dong2018a}). Even if such an instability were to develop, the resulting spiral arms are expected to be short lived and thus unlikely to be observed \citep{Hall2019}. Likewise, spirals resulting from varying shadows require a specific (and \textit{a priori} unlikely) synchronization of the shadow precession timescale with the orbital period at the launching point of the shadows \citep{Montesinos2016}. Therefore, both of these mechanisms appear unlikely to be the primary cause of spiral arms in circumstellar disks. 

On the other hand, the companion-driven hypothesis continues to be supported by a number of theoretical and observational lines of evidence (e.g., \citealt{Pohl2015}, \citealt{Dong2015b}, \citealt{Ren2018}). Recently, \cite{Dong2016} and \cite{Wagner2018b} used a combination of orbital monitoring, gas kinematics, and hydrodynamical simulations to show that the spiral arms around HD 100453 are driven by a low-mass stellar companion. This result represents the first confirmation of companion-driven spiral arms in a circumstellar disk, and provides further motivation for deep imaging of the other systems with spiral arms. 

For these other spiral-hosting systems (e.g., SAO 206462: \citealt{Muto2012}, MWC 758: \citealt{Grady2013}, LkH$\alpha$ 330: \citealt{Akiyama2016}, etc.), no similar companions have been reported. However, simulations predict that the companions could be of sufficiently low-mass to have evaded detection \citep{Dong2015b}. While recent observations have ruled out some of this parameter space (\citealt{Maire2017}, \citealt{Reggiani2018}), there is still an open possibility that planets on the lower end of this mass range ($\lesssim$3$-$5 M$_{Jup}$), or those that are less luminous than typical model predictions for a given mass (e.g., \citealt{Marley2007}, \citealt{Fortney2008}), are responsible for causing the observed spiral structures. 


\begin{deluxetable}{lcc}
\tabletypesize{\scriptsize}
\tablecaption{Properties of MWC 758}
\tablewidth{0pt}
\tablehead{\colhead{Parameter} & \colhead{Value} & \colhead{Ref.}}

\startdata
Mass & 1.5 $\pm$0.2 M$_{\odot}$ &  1 \\
Age & 3.5 $\pm$2.0 Myr & 2\\
Distance & 160.3 $\pm$1.7 pc & 3\\
$L^\prime$ & 4.75 & 4 \\
$M^\prime$ & 4.78 & 5
\enddata
\tablecomments{(1), \citep{Reggiani2018}, (2) \cite{Meeus2012}, (3) \cite{GAIADR2}, (4) \cite{Malfait1998}, (5) assuming $L^\prime$-$M^\prime$=$-0.03$ for an A8V star \citep{Mamajek2013}}
\end{deluxetable}

MWC 758 is an A8Ve star and likely member of the Taurus star-forming region. The general properties of the central star are listed in Table 1. Its protoplanetary disk has long been inferred from its infrared spectral energy distribution (SED), and recently garnered attention when it was discovered to host a pair of spiral arms in scattered light (\citealt{Grady2013}, \citealt{Benisty2015}). Simulations have suggested that the spiral arms could be caused by a planet at the end of one of the arms (\citealt{Dong2015b}, \citealt{Baruteau2019}), and that the azimuthal separation and contrast of the two main arms depend on the companion mass (\citealt{Fung2015}, \citealt{Dong2017}). Applying such dynamical mass constraints to MWC 758 yields a mass estimate of $M_{planet} \gtrsim$5 M$_{Jup}$. 

These predictions are broadly consistent with the observed rate of rotation of the spiral arms over the past decade \citep{Ren2018}, which indicates that the perturbing companion has an orbital period $\gtrsim$600 yr$-$placing it exterior to the spiral arms at a most likely separation of $\sim$0$\farcs$6. Furthermore, the Southern spiral arm in MWC 758 was found in ALMA mm continuum emission with a peak/background contrast of $\sim$3 (\citealt{Dong2018b}). This means it is a density structure and very unlikely to be a surface or shadow feature (further supporting the companion-driven hypothesis). Among the known disks with spiral arms, MWC 758 probably presents the best case for a planetary-origin argument. 

Our primary motivation behind observing MWC 758 is to detect the hypothetical substellar companion that is proposed to be responsible for the pair of spiral arms, or$-$in the absence of a detected companion$-$to improve existing detection limits.\footnote{In \S3, we show an achieved $\sim$2.5$\times$ sensitivity gain in the $L^\prime$ filter with respect to existing observations.} Given that circumstellar extinction may significantly reduce the apparent brightness of the hypothetically responsible companion, high-contrast observations at longer infrared wavelengths may reveal a companion that was previously un-identified at shorter wavelengths, which motivates our choice of the Large Binocular Telescope (LBT) due to its unique capabilities in the thermal infrared (see \S2). Observing in the thermal infrared is also motivated by recent work suggesting that young planets themselves may be obscured by circumplanetary material, which itself would radiate more strongly at mid-infrared wavelengths \citep{Szulagyi2019}. Finally, during our observing campaign a potential companion candidate was reported interior to the spiral arms (\citealt{Reggiani2018}, hereafter R18), and thus a secondary goal of our study became to attempt to re-image this companion candidate.


\section{Observations and Data Reduction}

The observations were carried out on 15 October 2016, 11 February 2017, and 25 December 2018 utilizing the LBT Mid-Infrared Camera (LMIRCam, \citealt{Leisenring2012}).  LMIRCam is located behind the cryogenic beam combiner of the LBT Interferometer (LBTI, \citealt{Hinz2016}).  The LBTI combines the light from the two 8.4m apertures for sensitive adaptive optics and interferometric high-contrast infrared observations (e.g., \citealt{Stone2018}, \citealt{Ertel2018}).  The atmospheric wavefront distortions are corrected using two adaptive secondary mirrors (one for each aperture) and high performance adaptive optics systems (\citealt{Esposito2010}, \citealt{Bailey2014}). 

We imaged the light from the two apertures simultaneously on separate areas of the LMIRCam detector.  This produces independent speckle patterns for the two apertures, significantly improving the fidelity of any circumstellar structures seen by both apertures.  A two-point, on-chip nod sequence was used for background subtraction. None of the observations utilized a coronagraph, which limits our sensitivity at small inner working angles, but yields comparable performance at angular separations exterior to the spiral arms. 


Due to technical issues with the adaptive optics system, only one aperture was in operation on 15 October 2016. Furthermore, during this period the images were impacted by diffraction from dust contamination on the Dewar window, which impacted the PSF in one of the dither positions. The instrument Dewar was cleaned before subsequent observations. The LBTI wavefront sensors were also upgraded in 2018$-$the new system (SOUL-AO; \citealt{Pinna2016}) provides a 1.7 kHz loop speed and 40x40 pupil subapertures compared to 1 kHz and 30x30 supapertures used previously. Finally, both apertures were in operation on 11 February 2017 and 25 December 2018. For these reasons, the data quality improved substantially with each successive observation.

We measured and subtracted the thermal background by taking the mean of the images within neighboring nod sequences, and then re-aligned the images via cross-correlation of the unsaturated Airy ring pattern, keeping independent datacubes for each dither position and each telescope aperture. Given that the PSF core is saturated in the majority of the images, we re-centered the images via a rotational-based centering algorithm (e.g., \citealt{Morzinski2015}). We measured the instrumental image distortion during the 2016 and 2017 epochs with an illuminated pinhole grid and corrected the images from each epoch independently using \texttt{Dewarp}  \citep{Spalding2019}. For the 2018 epoch, we take the $\lesssim$2-3 pixel uncertainty from the lack of distortion correction into consideration.

We subtracted the remaining background variations, including detector bias (primarily vertical striping), first by subtracting the mode of each column and row from the images, and, second, with a high-pass filter, which subtracts a smoothed version of the image. We then subtracted the PSF of the central star through a projection onto eigen-images (\citealt{Soummer2011}, specifically using the implementation presented in \citealt{Apai2016}). The properties of each data reduction are given at the bottom of  Table 2.

\begin{deluxetable}{lccc}
\tabletypesize{\scriptsize}
\tablecaption{Observing Log}
\tablewidth{0pt}
\tablehead{\colhead{Parameter} & \colhead{15-Oct-2016} & \colhead{11-Feb-2017} & \colhead{25-Dec-2018}  }

\startdata
Filter & $L^{\prime}$  & $L^{\prime}$  & $M^{\prime}$ \\
Avg. Seeing  & 0$\farcs$7  & 1$\farcs$0 & 1$\farcs$2 \\
Field Rot.  & 91$^\circ$ & 89$^\circ$ & 92$^\circ$ \\
Exp. Time & 0.87s & 0.87s & 0.151s\\
Total Int. Time  & 1.2 hr & 1.5 hr & 0.8 hr\\
\hline
\smallskip\\
\multicolumn{4}{c}{Data Reduction Parameters} \\
\hline\\
Frame Binning & 48 &  48 & 200\\
High Pass Width & 15 px & 15 px & 27 px\\
KLIP Components & 5 & 5 & 5 \\
Ann. Segments & 6 & 6 & 6 \\
Radial Range & 10-95 px & 10-95 px & 10-95 px\\ 
Ref. Angle Range& 1.2-20$^\circ$  & 1.2-89$^\circ$  & 1.5-92$^\circ$
\enddata
\tablecomments{Integration times are given as the sum of integration times for both apertures (as present for the 2017 and 2018 data) and excludes data taken in the rejected dither position in 2018.}
\end{deluxetable}




We refined the reduction parameters and assessed the quality of each datacube independently by injecting point sources into the raw data and then repeating the reduction. Finally, we discarded datacubes that were identified by eye to be significantly poorer quality than the others. This consisted of datacubes from the right aperture on the first night due to technical issues with the AO system, and one of the dither positions of the right aperture on the third night due to proximity with a damaged region of the detector. Finally, for each night we median-combined the images from the remaining frames into a master image.

\section{Results}

In Figure 1, we show our  $L^\prime$ and $M^\prime$ images obtained with the LBT taken on the nights of 11 February 2017, and 25 December 2018, respectively. We converted the images to signal to noise (S/N) maps by computing the signal measured in an aperture of FWHM diameter centered on each pixel, compared to the standard deviation of fluxes measured in non-overlapping apertures at the same stellocentric radius.\footnote{With the apertures positioned such that the first at each radius is at 0$^\circ$ N.} Specifically, we calculated S/N via Equation 9 in \cite{Mawet2014}, which accounts for small sample statistics at small angular separations.\footnote{Note that confidence intervals derived from these numbers are also a function of angular separation.} 

\begin{figure}[htpb]
\figurenum{1}
\epsscale{0.85}
\plotone{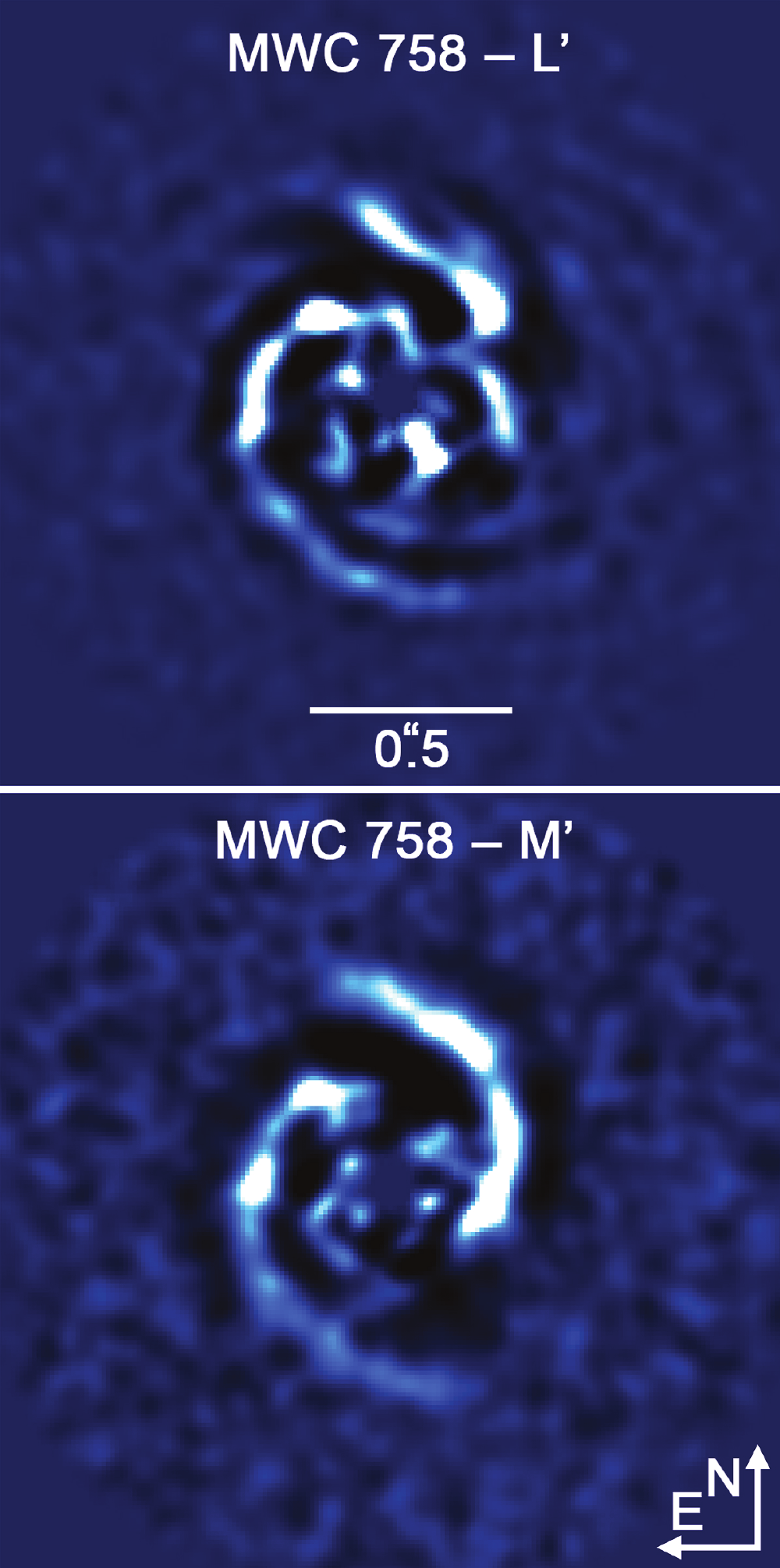}
\caption{\textbf{Top:}  $L^\prime$ image of MWC 758 taken on 11 February 2017. \textbf{Bottom:} $M^\prime$ image taken on 25 December 2018. Both datasets were taken in dual-aperture mode with the LBT and then combined. The images shown here are smoothed by a 5-pixel Gaussian kernel.}
\end{figure}

\begin{figure}[htpb]
\figurenum{2}
\epsscale{0.85}
\plotone{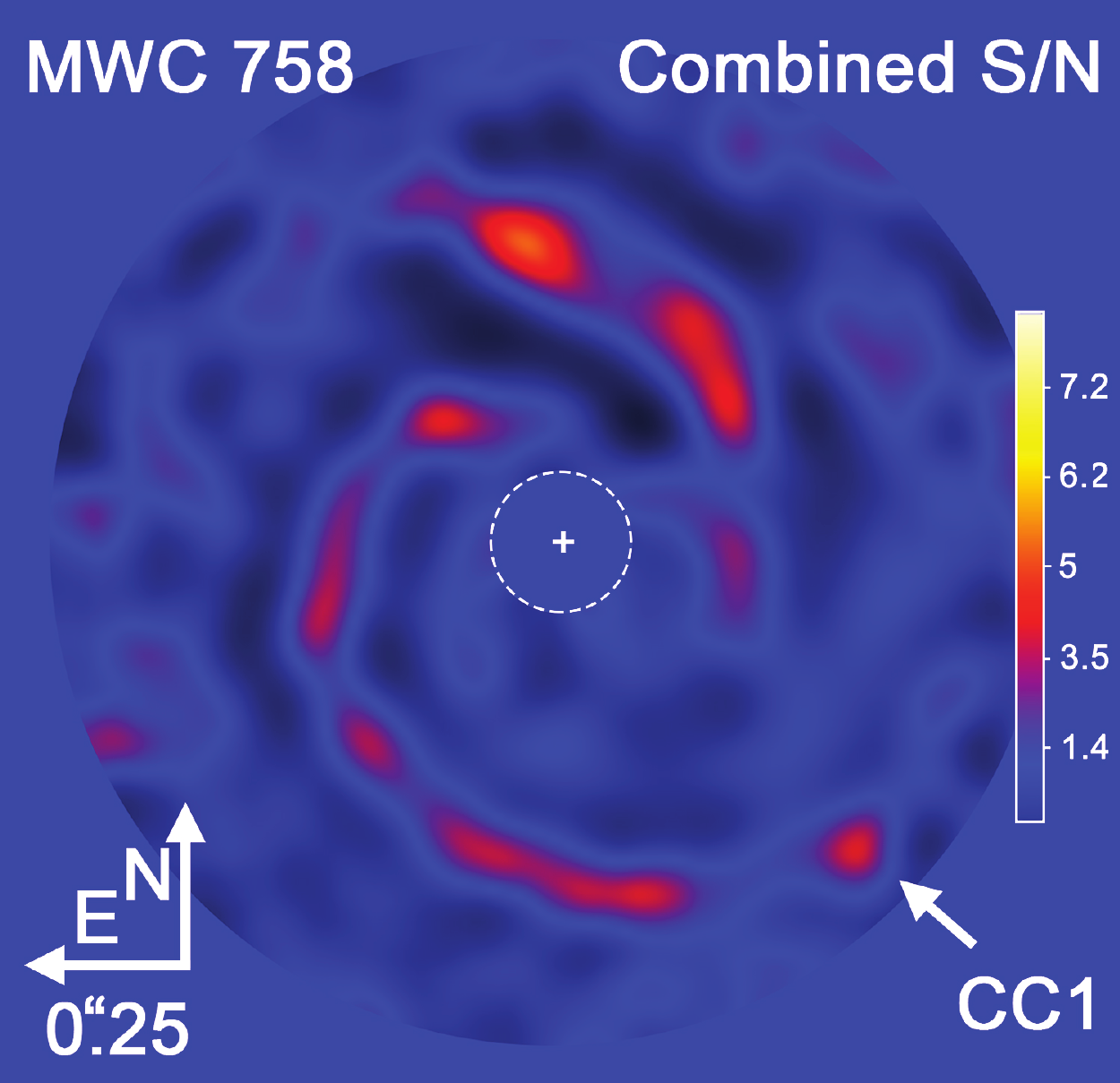}
\caption{Combined Signal to Noise (S/N) map of our three epochs of MWC 758 images taken with the LBT. A low-significance (S/N=3.9) source exterior to the Southern spiral arm is indicated as ``CC1".}
\end{figure}

\begin{figure*}[htpb]
\figurenum{3}
\epsscale{1.17}
\plotone{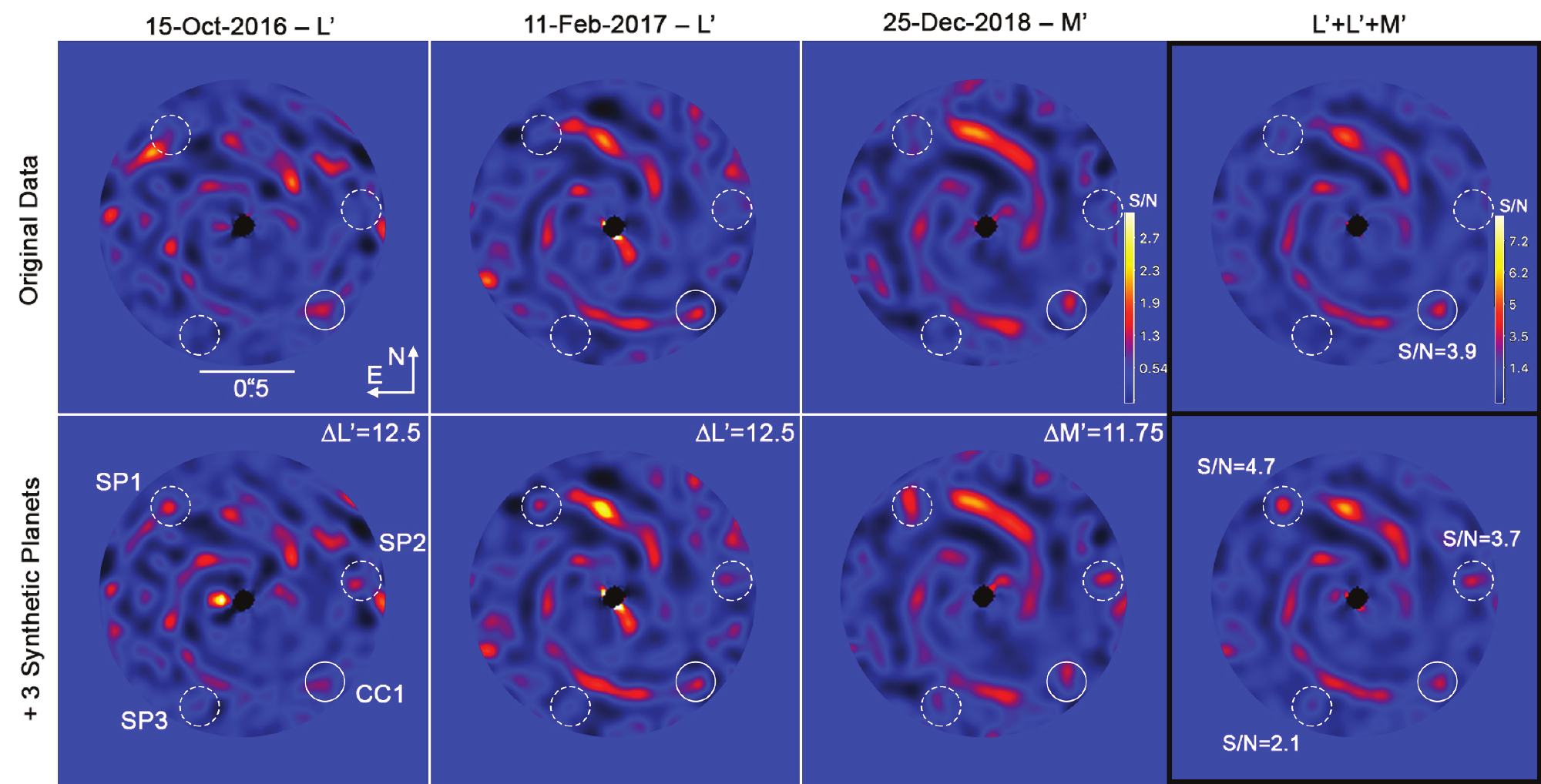}
\caption{Signal to Noise (S/N) maps of our three epochs of MWC 758 images taken with the LBT. The right-most column shows the total S/N maps, while the bottom row shows versions including three injected synthetic planets (SPs) to aid in image characterization. The injected sources, whose contrasts in the individual bands are listed in the image, are recovered with mean S/N$\sim$3.5. The original data show a tentative point source at the end of the Southern spiral arm with S/N$\sim$3.9 (3.6 when processed in the presence of the fake planets).}
\end{figure*}

In Figure 2, we show the combined S/N map, and in Figure 3, we show the S/N maps from each of the three epochs, their combination, and versions with synthetic planets injected to aid in the analysis of the images. At 0$\farcs$6 (exterior to the spiral arms), we recover sources with $\Delta L^\prime$=12.5 contrast with S/N=3.5. Following \cite{Stone2018}, we convert this value to one that can be directly compared to the 95\% completeness contrast curve reported in \cite{Reggiani2018}, which represents the most sensitive previously existing dataset. Exterior to the spiral arms, we find that our images are $\sim$2.5$\times$ more sensitive than those presented in \cite{Reggiani2018}.


\subsection{Disk Characterization}

Here, we characterize the observed properties of the disk, which is detected for the first time in the $M^\prime$ filter$-$probing deeper into the disk surface than previous infrared observations. We clearly detect two spiral arms in both the $L^\prime$ and $M^\prime$ images extending from $\sim 0\farcs2 - 0\farcs 5$. The contrast of the disk to the central star (and any unresolved inner disk emission) varies between $\sim$10$^{-4} - 10^{-5}$ per resolution element, depending on the filter and the location in each arm. Further characterization of bright/dark substructures within the arms is complicated by ADI-induced artifacts, that are most easily identifiable as dark spirals adjacent to the bright (physical) ones. 

The pitch angle of the spiral arms, and their angular separation, are useful properties to quantify as they have been suggested to be related to the mass of the disk and to the mass of the perturbing companion, respectively. While the structures observed in our images are in reality features in scattered light from the upper layers of the disk, for the sake of establishing simple estimates we will assume that the structures share similar properties to the mid-plane structures in the disk. We will also ignore the disk's small inclination from face-on. The pitch angle of the Northern arm varies from $\sim$26$^{\circ}-29^\circ$ from its base to end, while the pitch of the Southern arm varies from $\sim$25$^\circ -29^\circ$. The arms are separated by $\sim$142$^\circ$ at their base, $\sim 161^\circ$ at the end of the  Northern arm, and $\sim 171^\circ$ at the end of the (longer) Southern arm.  

\cite{Yu2019} recently found that pitch angle is tightly correlated with protoplanetary disk mass. Using the relationship between pitch angle and disk mass given in their Equation 2, we find log($M_{Disk}/M_{\odot})\lesssim -2.1\pm1.0$. This places MWC 758 in the regime in which it is at most marginally susceptible to gravitational instabilities, and in which a two-armed spiral pattern is unlikely to develop or be observed (\citealt{Kratter2016}, \citealt{Hall2019}). Meanwhile, \cite{Fung2015} used hydrodynamic and radiative transfer simulations to determine a relation between the separation of the spiral arms and the mass ratio between the perturbing companion to the central star, $q=\frac{M_{planet}}{M_{\star}}$. Using their Equation 9, we find $q$ = 0.005$-$0.014, corresponding to $M_{companion}$ =  7$-$25 M$_{Jup}$ for a stellar mass of $M_{\star}=1.5\pm0.2 M_{\odot}$.\footnote{This range includes the propagated uncertainty both on $q$ and on $M_{\star}$.} 

These constraints are not rigorous: we have assumed that the upper layers trace the structure at the mid-plane, and also that the disk is seen exactly face-on. However, these estimates establish a useful context: the disk is likely not massive enough to develop gravitational instabilities that would lead to a two-armed spiral structure, and, in that case, the companion responsible for driving the spiral arms is likely a giant planet or low-mass brown dwarf that could plausibly be detected in our images.

\subsection{Characterization of Potential Point-Sources}

Here, we characterize the properties of point sources in the close vicinity ($\leq$1") of MWC 758. Throughout this discussion, we will frequently refer to the combined S/N map in Figure 2, and to the individual S/N maps in Figure 3. To the SW of the disk, we detected a low S/N feature (hereafter referred to as Companion Candidate 1, or CC1) in a consistent position ($\rho=0\farcs 617\pm0\farcs024$, $\theta=$224.9$^\circ \pm2.2^\circ$ E of N) and contrast ($\Delta L^\prime = 12.5\pm0.5$) in both $L^\prime$ images. The $M^\prime$ image shows a source in the same location and slightly brighter ($\Delta M^\prime =  11.75\pm0.5$). The source adds constructively in the combination of the S/N maps and appears with a final significance of S/N=3.9.

The red color of CC1 and its position at the end of the spiral arm are consistent with a planetary nature of the source,\footnote{With most of the flux emanating from either a planetary photosphere, or from a circumplanetary disk.} and thus we convert its luminosity to mass-estimates of 2 $-$ 5 $M_{Jup}$ via the ``hot-start" COND models (\citealt{Baraffe2003}, \citealt{Baraffe2015}), and $\lesssim$13 $M_{Jup}$ if we account for planets that possibly formed with less than maximal luminosity (e.g., \citealt{Fortney2008}). These estimates assume negligible extinction and emission from circumplanetary  and/or circumstellar material. The inferred mass would be larger in the case of significant extinction, and perhaps significantly different than these estimates in the case that emission from circumplanetary material is the dominant signal (e.g., \citealt{Szulagyi2019}).

In \S4.1.1, we will discuss the probability that CC1 is a real (astrophysical) detection vs. a false positive (a residual atmospheric or instrumental speckle). In \S4.1.2, we will discuss potential astrophysical natures of the source$-$namely either emission from a protoplanet and/or circumplantery disk, or point-like emission from a component of the circumstellar disk.

\subsection{Limits on Planets Exterior to the Spiral Arms}


It is useful to know what masses of planets are compatible with being undetectable in the data. We assessed our detection limits based on synthetic planet injections, and found that we could identify planets with S/N=5 down to $\sim$1.5$\times$10$^{-5}$ contrast to the host star in $L^\prime$, and $\sim$3$\times$10$^{-5}$ in $M^\prime$. Combined with the star's age, distance, and photometry (see Table 1), we convert this contrast limit into an upper mass limit of 3 $-$ 5 M$_{Jup}$ (via the models of \citealt{Baraffe2003} and \citealt{Baraffe2015}) for planets exterior to the spiral arms.\footnote{This estimated mass range includes propagated uncertainty on age, distance, and photometry of the companion and host star.} Using the cold-start prescription in \cite{Wagner2019}, we similarly convert our photometric sensitivity to an upper mass limit for cold-start planets of 5 $-$ 13 M$_{Jup}$, with the larger uncertainty reflecting the uncertainty in initial planetary entropy post-hydrodynamic collapse (assuming a uniform distribution between 50 $-$ 100\% of initial conserved entropy). The upper end of the latter mass range is consistent with masses required to drive spiral arms in hydrodynamical simulations (e.g., \citealt{Dong2015b}).



\subsection{Limits on Companions Interior to the Spiral Arms}



A recent study on MWC 758 by \cite{Reggiani2018} reported several new and interesting features, including a companion candidate at $\sim$0$\farcs$11 from the central star and contrast of $\Delta L^\prime \sim$ 7.0. The companion candidate was recovered on two separate nights in 2015 and 2016 at consistent position and brightness, leading the discovery team to posit that the source is an orbiting companion with a period of $\sim$60 years. Such a companion would have orbited $\sim$18$^\circ$ in the clockwise direction between the 2015 measurement in \cite{Reggiani2018} and our $M^\prime$ observation on 25 December 2018, which provides the most extended time baseline. 

\begin{figure}[htpb]
\figurenum{4}
\epsscale{1.1}
\plotone{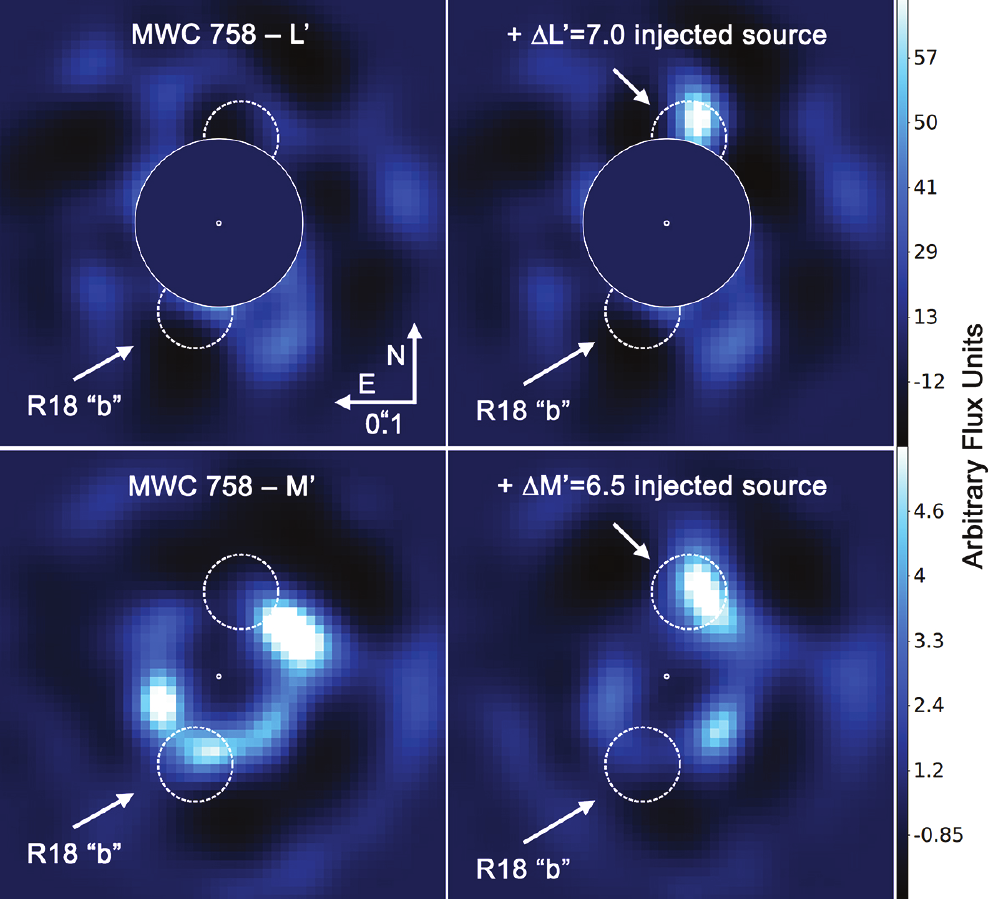}
\caption{The inner regions of MWC 758, with the location of the candidate companion reported in \cite{Reggiani2018} (R18) indicated by the lower dashed circle to the SSE of the star, whose position is marked at the center of each image. We simulated the anticipated detection of the candidate by injecting a synthetic point source (upper dashed circle) into the raw data with the reported contrast and angular separation, but at opposite PA, and recovered the injected source with S/N$\sim$4. The intensity scales are identical between left and right images, and the differences in the residual speckle patterns are due to the injected source entering into the KLIP processing. The solid blue circle at the center of the $L^\prime$ images masks the saturated region of the detector. The images shown here are smoothed by a 5-pixel Gaussian kernel.}
\end{figure}

To test whether we could recover the candidate reported by \cite{Reggiani2018}, we first considered whether such a detection would be plausible given that our images are saturated near the center of MWC 758. We verified that the $M^\prime$ data are not saturated beyond $\sim$2-3 pixels from the center of MWC 758, enabling accurate image characterization through synthetic companion injections at 0$\farcs$11, or 11 pixels from image center. However, in $L^\prime$, the detector was saturated out to $\sim$10 pixels from the image center, while the FWHM region of the hypothetical candidate's PSF would extend from $\sim$6-15 pixels, causing it to be partially cut off near the star due to overlapping with the saturated region. Thus, the $M^\prime$ data provide the greatest potential for re-detecting the candidate presented by \cite{Reggiani2018}, especially given that protoplanet and circumplanetary disk luminosities are expected to increase from $L^\prime$ to $M^\prime$ (\citealt{Eisner2015}, \citealt{Zhu2015}). The $L^\prime$ data are still useful to probe companions near to the central star, with the caveat that the PSF of any potential sources will appear to be cut-off within the saturated region of the detector ($\lesssim$0$\farcs$1). 

We reduced our three epochs of LBT data again in a similar manner to the reduction described in \S2, with several alterations to the previous strategy to improve performance at small inner working angles. These changes included a decrease in the extent of the KLIP-reduced region to 5-20 pixels (vs. 10-95 pixels), and processing the data in full annuli, rather than annular segments. The latter change was made in order to increase the area of the optimization region in response to the decrease in area imposed by the former change. We also increased the number of KLIP components from 5 to 10, and increased the minimum accepted angular separation for reference PSFs to 4$^\circ$ and 5$^\circ$ for the $L^\prime$ the $M^\prime$ datasets, respectively, in order to mitigate self-subtraction. Additionally, we injected a synthetic point source at the same separation as the reported candidate, but 180$^\circ$ opposite in PA. This synthetic source was injected at the reported contrast of $\Delta L^\prime$=7.0, and 0.5 mag brighter in $M^\prime$ to (conservatively) account for the color difference expected for emission from a circumplanetary accretion disk (\citealt{Eisner2015}, \citealt{Zhu2015}).

In Figure 4, we show the $L^\prime$ and $M^\prime$ images of the inner regions of MWC 758. We do not detect a source near the reported location and brightness of the candidate in \cite{Reggiani2018}. However, we do recover the injected planet with S/N$\sim$4, suggesting that the candidate would have likely been detected in our observations if it were an astrophysical source.\footnote{Sources injected at different position angles, including the reported location of the candidate itself, are also recovered with S/N$\sim$4.} At the location of the putative source, we measure S/N=1.5. Following \cite{Mawet2014}, we utilized a t-distribution with a shape parameter appropriate for the 0$\farcs$11 separation to calculate a $\sim$3\% chance of observing a real $\Delta L^\prime$=7.0 source with S/N=1.5. That is, there is a $\sim$3\% chance that the reported source is overlaid by a negative speckle with a significance of $\sim$2.5$\times$ the standard noise level. This non-detection is also consistent with a recent non-detection in H$\alpha$, which suggests that if the source is indeed a companion, that it lacks a strong accretion signature (\citealt{Huelamo2018}). We therefore suggest that this source be considered as a possible spurious detection, and that the designation ``MWC 758b" should be reserved for when the first companion to the system is definitively discovered.

\section{Discussion}

\subsection{A Planet at the End of the Southern Arm?}

An intriguing result of our observations is the tentative detection of a point source at the end of the Southern arm (CC1). Forward modelling of synthetic point sources at the same angular separation suggests that the feature is $\sim$10$^{-5}$ contrast in $L^\prime$, and $\sim$2$\times$10$^{-5}$ contrast in $M^\prime$, which translate to a mass-estimate of $\sim$2 $-$ 5 M$_{Jup}$ via the models of \cite{Baraffe2003} and \cite{Baraffe2015}, assuming a photospheric nature of the emission. Prior to discussing the potential nature of this feature, we first discuss the probability that it represents a real detection of an astrophysical object (namely, a planet or part of a disk), or whether it is a spurious signal in the random speckle pattern.

\subsubsection{CC1 False Positive Probability}

Here, we consider the probability that CC1 is a false positive (i.e., a residual speckle of atmospheric or instrumental origin) with consistent brightness and position in each of our three images. In an area exterior to the spiral arms ($\geq$0$\farcs$5) and extending radially to just exterior to CC1's position ($\leq$0$\farcs$75), we count the number of speckles with a similar positive S/N to CC1 (including CC1 itself) on each of the three nights. This essentially constitutes all sources (other than the known spiral arms) that have a red color in Figure 2. We arrived at $\sim$10 such sources for each epoch, or $\sim$10 false positive sources arcsec$^{-2}$. The possibility that a false positive detection that occurred randomly on one night would occur with consistent brightness and position (within a circular aperture with Diameter=2$\times$FWHM, or area = 0.027 arcsec$^2$) on a different night is thus $\sim$30\%, and on yet another night is $\sim$9\%. Thus, we consider that CC1 is likely of astrophysical origin with approximately 90\% confidence. While there still exists a significant ($\sim$10\%) possibility that the source is a false positive, we will next briefly consider some plausible natures of this source, under the hypothesis that it is a true detection.

\subsubsection{Possible Physical Natures of CC1}
CC1 lies exterior to the spiral arms detected in scattered light (\citealt{Grady2013}, \citealt{Benisty2015}, \citealt{Reggiani2018}), but nevertheless could likely be explained by emission (or scattering) from deeper within the disk. Alternatively, the feature could be emission from a planetary photosphere or circumplanetary disk (CPD). The latter scenario is tentatively supported by a number of arguments: 1) the $L^\prime-M^\prime$ color of the source is positive, which is consistent with emission from a planetary photosphere or CPD (\citealt{Eisner2015}, \citealt{Zhu2015}); 2) tracking of the rotation of the spiral pattern over the past decade suggests that the perturbing companion should be at $\sim$0$\farcs$6 \citep{Ren2018}, which is consistent with the measured separation of CC1; and 3) hydrodynamical simulations predict that the planet responsible for the arms should be more massive than $\sim$5 M$_{Jup}$, which translates into $\Delta L^\prime \leq11.5$ and $\Delta M^\prime \leq11.0$ for hot-start models and no extinction, and somewhat fainter for less energetic initial conditions or cases of high extinction. In general, these predictions are similar to the measured properties of CC1 and support the explanation that CC1 is the spiral arm driving planet. 

While we cannot presently rule out the possibility that CC1 may be an unresolved feature in the circumstellar disk, such a possibility appears to be qualitatively ruled out given that the rest of the disk exhibits relatively gray scattering. The contrasts (per resolution element) in $L^\prime$, $M^\prime$, and at shorter wavelengths (such as $H-$ and $K$-bands) are relatively similar at $\Delta (H,K,L^\prime, M^\prime)\lesssim$ 12.5. Meanwhile, \cite{Grady2013} established a detection limit of $\Delta K=12.5$ at the position of CC1. Furthermore, \cite{Benisty2015} did not detect a polarized light source at $\sim$1 $\mu m$ in the vicinity of CC1, which provides additional evidence that CC1 is not a circumstellar disk feature.

Therefore, the lack of a detection of CC1 at shorter wavelengths is perhaps the strongest evidence supporting the planetary nature of CC1, similar to the protoplanet candidate HD 100546b (\citealt{Rameau2017}, \citealt{Follette2017}, \citealt{Currie2017}). Overall, we consider CC1 as a viable candidate for the planet driving the spiral arms around MWC 758. However, given the challenging interpretation of the emission and $\sim$10\% probability of the source being a false positive detection, we defer further classification and characterization of this source for future, more sensitive observations. Improved data in the thermal infrared and high-contrast H$\alpha$ observations may help to discern whether CC1 is in fact the spiral-arm driving planet in MWC 758. 

Future astrometric monitoring will also be useful to confirm/reject whether CC1 is of planetary nature. Considering the source's projected separation combined with the distance and mass of the host star, we estimate an orbital period of $\sim$800 yr, which would generate $\sim$0.5$^\circ yr^{-1}$ of orbital motion. This anticipated motion of CC1 should be resolvable in a few years with current instrumentation. Furthermore, the spiral pattern in the disk should rotate with a similar angular frequency. Based on ten years of disk imaging, \cite{Ren2018} found an angular velocity for the arms of 0.6$^\circ$ $_{-0.6^\circ}^{+3.3^\circ} yr^{-1}$. Improved constraints on the rotation rate of the spiral arms could strengthen the link between the spiral arms and CC1 as the driving companion.

\subsection{Two vs. Three Spiral Arms?}

\cite{Reggiani2018} also reported the detection of a third spiral arm between the two spiral arms that had been reported previously by \cite{Grady2013} and later by \cite{Benisty2015}. From a theoretical standpoint, such a feature is unlikely to represent an actual third arm (most simulations of spiral density waves in protoplanetary disks result in two prominent arms).  However, an apparent third arm could possibly represent the back-side of the disk, as has been observed for other disks (e.g., HD 100453: \citealt{Benisty2017}). While there is some residual flux in this region of our $L^\prime$ data (when examined in raw image counts), the S/N maps do not show any significant feature (S/N$>$3) in this location, despite the fact that our data probe similar angular scales and contrasts. Furthermore, there is no similar feature in the $M^\prime$ data, which probe the deepest layers of the disk among the images considered here and in \cite{Reggiani2018}. We therefore suggest that the third arm identified by \cite{Reggiani2018} is likely a data processing artifact, and does not trace a physical disk structure.

\section{Summary and Conclusions}

1) We presented new images of the disk around MWC 758 taken between October 2016 and December 2018, constituting the deepest yet in $L^\prime$ and the first images recorded of the system at $M^\prime$.

2) We identified a low significance-level source (CC1) located near the end of the Southern arm that is detected on each of our three nights, and discussed three possibilities for CC1: a) a non-astrophysical data artifact, b) an unresolved circumstellar disk feature, and c) a giant planet that is possibly related to the spiral arms.

3) We established a 5$\sigma$ contrast limit on companions exterior to the spiral arms ($\gtrsim$0$\farcs5$) of $\sim$1.5$\times$10$^{-5}$ contrast to the host star (and its unresolved inner disk) in $L^\prime$, and $\sim$3$\times$10$^{-5}$ in $M^\prime$.

4) We converted these contrast limits into upper mass limits via hot$-$ and cold-start evolutionary models and arrived at upper masses of planets that could exist exterior to the spiral arms to be $\sim$3$-$4 M$_{Jup}$ and 4$-$13 M$_{Jup}$ for hot$-$ and cold-start planets, respectively, assuming M$_{L^\prime}$=4.75 for the central star and negligible extinction and emission from circumplanetary and circumstellar material. 

5) We did not detect the companion candidate recently reported by \cite{Reggiani2018}, although forward modelling of the anticipated signal suggests that we would have been able to detect a source with the reported characteristics.




\section{Acknowledgments}

The authors express their sincere gratitude to LBT Director Christian Veillet for allocating director's time for this project, to Maddalena Reggiani, Thayne Currie, and the anonymous referee for reviewing and providing feedback on earlier versions of this manuscript, and to Jenny Power, Amali Vaz, Lauren Schatz, and Carl Coker who assisted with the observations. The results reported herein benefited from collaborations and/or information exchange within NASA's Nexus for Exoplanet System Science (NExSS) research coordination network sponsored by NASA's Science Mission Directorate. J.M.S. is supported by NASA through Hubble Fellowship grant HST-HF2-51398.001-A awarded by the Space Telescope Science Institute, which is operated by the Association of Universities for Research in Astronomy, Inc., for NASA, under contract NAS5-26555. The LBT is an international collaboration among institutions in the United States, Italy, and Germany. LBT Corporation partners are: The University of Arizona on behalf of the Arizona university system; Istituto Nazionale di Astrofisica, Italy; LBT Beteiligungsgesellschaft, Germany, representing the Max-Planck Society, the Astrophysical Institute Potsdam, and Heidelberg University; The Ohio State University, and The Research Corporation, on behalf of The University of Notre Dame, University of Minnesota, and University of Virginia.






\begin{thebibliography}{}

\bibitem[Akiyama et al.(2016)]{Akiyama2016} Akiyama, E., Hashimoto, J., Liu, H.~B., et al.\ 2016, \aj, 152, 222 

\bibitem[Andrews et al.(2018)]{Andrews2018} Andrews, S.~M., Huang, J., P{\'e}rez, L.~M., et al.\ 2018, \apjl, 869, L41 

\bibitem[Apai et al.(2015)]{Apai2015} Apai, D., Schneider, G., Grady, C.~A., et al.\ 2015, \apj, 800, 136 

\bibitem[Apai et al.(2016)]{Apai2016} Apai, D., Kasper, M., Skemer, A., et al.\ 2016, \apj, 820, 40 

\bibitem[Bailey et al.(2014)]{Bailey2014} Bailey, V.~P., Hinz, P.~M., Puglisi, A.~T., et al.\ 2014, \procspie, 9148, 914803 

\bibitem[Baraffe et al.(2003)]{Baraffe2003} Baraffe, I., Chabrier, G., Barman, T.~S., Allard, F., \& Hauschildt, P.~H.\ 2003, \aap, 402, 701 

\bibitem[Baraffe et al.(2015)]{Baraffe2015} Baraffe, I., Homeier, D., Allard, F., \& Chabrier, G.\ 2015, \aap, 577, A42 

\bibitem[Baruteau et al.(2019)]{Baruteau2019} Baruteau, C., Barraza, M., P{\'e}rez, S., et al.\ 2019, \mnras, 486, 304 

\bibitem[Benisty et al.(2015)]{Benisty2015} Benisty, M., Juhasz, A., Boccaletti, A., et al.\ 2015, \aap, 578, L6 

\bibitem[Benisty et al.(2017)]{Benisty2017} Benisty, M., Stolker, T., Pohl, A., et al.\ 2017, \aap, 597, A42 


\bibitem[Currie et al.(2017)]{Currie2017} Currie, T., Brittain, S., Grady, C.~A., Kenyon, S.~J., \& Muto, T.\ 2017, Research Notes of the American Astronomical Society, 1, 40 

\bibitem[Debes et al.(2016)]{Debes2016} Debes, J.~H., Jang-Condell, H., \& Schneider, G.\ 2016, \apjl, 819, L1 

\bibitem[Dong et al.(2015a)]{Dong2015a} Dong, R., Zhu, Z., \& Whitney, B.\ 2015, \apj, 809, 93 

\bibitem[Dong et al.(2015b)]{Dong2015b} Dong, R., Zhu, Z., Rafikov, R.~R., \& Stone, J.~M.\ 2015, \apjl, 809, L5 

\bibitem[Dong et al.(2016)]{Dong2016} Dong, R., Zhu, Z., Fung, J., et al.\ 2016, \apjl, 816, L12 

\bibitem[Dong \& Fung(2017)]{Dong2017} Dong, R., \& Fung, J.\ 2017, \apj, 835, 38 

\bibitem[Dong et al.(2018a)]{Dong2018a} Dong, R., Najita, J.~R., \& Brittain, S.\ 2018, \apj, 862, 103 

\bibitem[Dong et al.(2018b)]{Dong2018b} Dong, R., Liu, S.-y., Eisner, J., et al.\ 2018, \apj, 860, 124 

\bibitem[Eisner(2015)]{Eisner2015} Eisner, J.~A.\ 2015, \apjl, 803, L4 

\bibitem[Ertel et al.(2018)]{Ertel2018} Ertel, S., Defr{\`e}re, D., Hinz, P., et al.\ 2018, \aj, 155, 194 

\bibitem[Esposito et al.(2010)]{Esposito2010} Esposito, S., Riccardi, A., Fini, L., et al.\ 2010, \procspie, 7736, 773609 

\bibitem[Follette et al.(2017)]{Follette2017} Follette, K.~B., Rameau, J., Dong, R., et al.\ 2017, \aj, 153, 264 

\bibitem[Fortney et al.(2008)]{Fortney2008} Fortney, J.~J., Marley, M.~S., Saumon, D., \& Lodders, K.\ 2008, \apj, 683, 1104 

\bibitem[Fung \& Dong(2015)]{Fung2015} Fung, J., \& Dong, R.\ 2015, \apjl, 815, L21 

\bibitem[Gaia Collaboration et al.(2018)]{GAIADR2} Gaia Collaboration, Brown, A.~G.~A., Vallenari, A., et al.\ 2018, \aap, 616, A1
\bibitem[Grady et al.(2013)]{Grady2013} Grady, C.~A., Muto, T., Hashimoto, J., et al.\ 2013, \apj, 762, 48 

\bibitem[Hall et al.(2019)]{Hall2019} Hall, C., Dong, R., Rice, K., et al.\ 2019, \apj, 871, 228 

\bibitem[Hinz et al.(2016)]{Hinz2016} Hinz, P.~M., Defr{\`e}re, D., Skemer, A., et al.\ 2016, \procspie, 9907, 990704 

\bibitem[Hu{\'e}lamo et al.(2018)]{Huelamo2018} Hu{\'e}lamo, N., Chauvin, G., Schmid, H.~M., et al.\ 2018, \aap, 613, L5 

\bibitem[Keppler et al.(2018)]{Keppler2018} Keppler, M., Benisty, M., M{\"u}ller, A., et al.\ 2018, \aap, 617, A44 

\bibitem[Kratter \& Lodato(2016)]{Kratter2016} Kratter, K., \& Lodato, G.\ 2016, \araa, 54, 271 


\bibitem[Leisenring et al.(2012)]{Leisenring2012} Leisenring, J.~M., Skrutskie, M.~F., Hinz, P.~M., et al.\ 2012, \procspie, 8446, 84464F 



\bibitem[Maire et al.(2017)]{Maire2017} Maire, A.-L., Stolker, T., Messina, S., et al.\ 2017, \aap, 601, A134 


\bibitem[Malfait et al.(1998)]{Malfait1998} Malfait, K., Bogaert, E., \& Waelkens, C.\ 1998, \aap, 331, 211 

\bibitem[Marley et al.(2007)]{Marley2007} Marley, M.~S., Fortney, J.~J., Hubickyj, O., Bodenheimer, P., \& Lissauer, J.~J.\ 2007, \apj, 655, 541 


\bibitem[Mawet et al.(2014)]{Mawet2014} Mawet, D., Milli, J., Wahhaj, Z., et al.\ 2014, \apj, 792, 97 

\bibitem[Meeus et al.(2012)]{Meeus2012} Meeus, G., Montesinos, B., Mendigut{\'{\i}}a, I., et al.\ 2012, \aap, 544, A78 

\bibitem[Montesinos et al.(2016)]{Montesinos2016} Montesinos, M., Perez, S., Casassus, S., et al.\ 2016, \apjl, 823, L8 

\bibitem[Morzinski et al.(2015)]{Morzinski2015} Morzinski, K.~M., Males, J.~R., Skemer, A.~J., et al.\ 2015, \apj, 815, 108 

\bibitem[Muley et al.(2019)]{Muley2019} Muley, D., Fung, J., \& van der Marel, N.\ 2019, \apjl, 879, L2 

\bibitem[Muto et al.(2012)]{Muto2012} Muto, T., Grady, C.~A., Hashimoto, J., et al.\ 2012, \apjl, 748, L22 

\bibitem[Pecaut \& Mamajek(2013)]{Mamajek2013} Pecaut, M.~J., \& Mamajek, E.~E.\ 2013, \apjs, 208, 9 

\bibitem[Pinna et al.(2016)]{Pinna2016} Pinna, E., Esposito, S., Hinz, P., et al.\ 2016, \procspie, 9909, 99093V 

\bibitem[Pohl et al.(2015)]{Pohl2015} Pohl, A., Pinilla, P., Benisty, M., et al.\ 2015, \mnras, 453, 1768 

\bibitem[Quanz et al.(2013a)]{Quanz2013a} Quanz, S.~P., Avenhaus, H., Buenzli, E., et al.\ 2013, \apjl, 766, L2 

\bibitem[Quanz et al.(2013b)]{Quanz2013b} Quanz, S.~P., Amara, A., Meyer, M.~R., et al.\ 2013, \apjl, 766, L1 

\bibitem[Rameau et al.(2017)]{Rameau2017} Rameau, J., Follette, K.~B., Pueyo, L., et al.\ 2017, \aj, 153, 244 

\bibitem[Reggiani et al.(2018)]{Reggiani2018} Reggiani, M., Christiaens, V., Absil, O., et al.\ 2018, \aap, 611, A74 

\bibitem[Ren et al.(2018)]{Ren2018} Ren, B., Dong, R., Esposito, T.~M., et al.\ 2018, \apjl, 857, L9 


\bibitem[Soummer et al.(2012)]{Soummer2011} Soummer, R., Pueyo, L., \& Larkin, J.\ 2012, \apjl, 755, L28 


\bibitem[Spalding(2019)]{Spalding2019} Spalding, E., \& Stone, J., \ 2019, Astrophysics Source Code Library, 1907.008

\bibitem[Stone et al.(2018)]{Stone2018} Stone, J.~M., Skemer, A.~J., Hinz, P.~M., et al.\ 2018, \aj, 156, 286 

\bibitem[Szul{\'a}gyi et al.(2019)]{Szulagyi2019} Szul{\'a}gyi, J., Dullemond, C.~P., Pohl, A., \& Quanz, S.~P.\ 2019, \mnras, 487, 1248 

\bibitem[van der Marel et al.(2019)]{vanderMarel2019} van der Marel, N., Dong, R., di Francesco, J., Williams, J.~P., \& Tobin, J.\ 2019, \apj, 872, 112 


\bibitem[Wagner et al.(2018a)]{Wagner2018a} Wagner, K., Follete, K.~B., Close, L.~M., et al.\ 2018, \apjl, 863, L8 

\bibitem[Wagner et al.(2018b)]{Wagner2018b} Wagner, K., Dong, R., Sheehan, P., et al.\ 2018, \apj, 854, 130 

\bibitem[Wagner et al.(2019)]{Wagner2019} Wagner, K., Apai, D., \& Kratter, K.~M.\ 2019, \apj, 877, 46 

\bibitem[Yu et al.(2019)]{Yu2019} Yu, S.-Y., Ho, L.~C., \& Zhu, Z.\ 2019, \apj, 877, 100 

\bibitem[Zhang et al.(2018)]{Zhang2018} Zhang, S., Zhu, Z., Huang, J., et al.\ 2018, \apjl, 869, L47 

\bibitem[Zhu et al.(2012)]{Zhu2012} Zhu, Z., Nelson, R.~P., Dong, R., Espaillat, C., \& Hartmann, L.\ 2012, \apj, 755, 6 

\bibitem[Zhu(2015)]{Zhu2015} Zhu, Z.\ 2015, \apj, 799, 16 




\end{thebibliography}
\end{document}